\documentclass{article}
\usepackage{amsmath,amssymb,amsfonts}   
\usepackage{graphicx,bm}

\usepackage{graphicx,bm}
  \usepackage{paralist}
  \usepackage{epstopdf}
  \usepackage{graphics} 
 \usepackage[colorlinks=true]{hyperref}
 \hypersetup{urlcolor=blue, citecolor=red}
 \usepackage[latin1]{inputenc}
\usepackage[caption=false]{subfig}

\renewcommand{\theequation}{\arabic{section}.\arabic{equation}}
 
 \renewcommand{\u}{\widetilde{u}}

\newcommand{\e}{{\rm e}}

\renewcommand{\P}{\mathbb{P}}

\newcommand{\E}{\mathbb{E}}

\newcommand{\U}{\widetilde{U}}
\newcommand{\V}{\widetilde{V}}

\def\P{\mathbb{P}}

\newcommand{\Markov}[2]{\underset{#1}{\overset{#2}{\rightleftharpoons}}}

\begin{document}

\title{Local accumulation time for diffusion in cells with gap junction coupling}

\author{ \em
P. C. Bressloff, \\ Department of Mathematics, 
University of Utah \\155 South 1400 East, Salt Lake City, UT 84112}

 \maketitle

\begin{abstract}

In this paper we analyze the relaxation to steady-state of intracellular diffusion in a pair of cells with gap-junction coupling. Gap junctions are prevalent in most animal organs and tissues, providing a direct diffusion pathway for both electrical and chemical communication between cells. Most analytical models of gap junctions focus on the steady-state diffusive flux and the associated effective diffusivity. Here we investigate the relaxation to steady state in terms of the so-called local accumulation time. The latter is commonly used to estimate the time to form a protein concentration gradient during morphogenesis. The basic idea is to treat the fractional deviation from the steady-state concentration as a cumulative distribution for the local accumulation time. One of the useful features of the local accumulation time is that it takes into account the fact that different spatial regions can relax at different rates. We consider both static and dynamic gap junction models. The former treats the gap junction as a resistive channel with effective permeability $\mu$, whereas the latter represents the gap junction as a stochastic gate that randomly switches between an open and closed state. The local accumulation time is calculated by solving the diffusion equation in Laplace space and then taking the small-$s$ limit. We show that the accumulation time is a monotonically increasing function of spatial position, with a jump discontinuity at the gap junction. This discontinuity vanishes in the limit $\mu \rightarrow \infty$ for a static junction and $\beta \rightarrow 0$ for a stochastically-gated junction, where $\beta$ is the rate at which the gate closes.
 Finally, our results are generalized to the case of a linear array of cells with nearest neighbor gap junction coupling.

\end{abstract}

\maketitle

\section{Introduction}

Gap junctions are small nonselective channels that provide a direct diffusion pathway between neighboring cells. They are formed by the head-to-head connection of two hemichannels or connexons, one from each of the two coupled cells  \cite{Evans02,Saez03,Good09}. Gap junctions are prevalent in most animal organs and tissues, providing a mechanism for both electrical and chemical communication between cells. Electrical coupling is particularly important in cardiac muscle, where the efficient transmission of electrical signals allows the heart muscle cells to contract in unison. Gap junctions or electrical synapses are also found throughout the central nervous system \cite{Connors04}. Direct chemical communication between cells occurs through the transmission of small second messengers, such as inositol triphosphate (IP3) and calcium (Ca$^{2+}$). An important example of long-range chemical signaling mediated by gap junctions is the propagation of intercellular Ca$^{2+}$ waves \cite{Sanderson12}.

One of the characteristic properties of a gap junction is its effective channel permeability $\mu$. Mathematically speaking, the high resistance to diffusive flow generates a jump discontinuity $\Delta u$ of molecular concentration across the gap junction such that the diffusive flux through the channel is given by $J=-\mu \Delta u$. A number of studies have analyzed the diffusion equation for a one-dimensional (1D) array of cells with nearest-neighbor coupling \cite{Brink85,Ramanan90,Keener09}. In particular, one can calculate the steady-state concentration in each cell by assuming that there is a constant diffusive flux $J_0$ through the cellular array; the flux is then determined self-consistently by solving the resulting boundary value problem. The latter includes a set of interior boundary conditions that combine jump discontinuities in the concentration at the gap junctions with flux continuity conditions. Analogous to the opening and closing of ion channels \cite{Smith02}, gap junctions can be both voltage-gated and chemically-gated \cite{Verselis04,Paul09}. This has motivated a stochastic model in which each gap junction is treated as a randomly fluctuating gate that switches between an open and a closed state \cite{Bressloff16}. Solving the resulting first-order moment equations for the stochastic concentrations and fluxes in steady state generates an effective channel permeability that depends on the kinetics of the stochastic gate, even though the channel is fully conducting when in the open state. 

One advantage of analytical models is that they provide explicit expressions for the steady-state flux $J_0$. The latter can be used to extract the effective diffusivity $D_e$ for the intracellular transfer of molecules via gap junctions, which can then be compared with experimental data. However, $D_e$ is a lumped parameter that depends on both the cytoplasmic diffusivity $D$ and the junctional permeability $\mu$. In order to separate these two biophysical parameters, it is necessary to include additional information about the diffusion process, such as the time-dependent approach to steady-state \cite{Ramanan90,Keener09}.

In this paper, we extend previous studies of static and dynamic gap junctions by investigating the relaxation to steady state in terms of the so-called local accumulation time. The latter is commonly used to estimate the time to form a protein concentration gradient during morphogenesis \cite{Berez10,Berez11,Berez11a,Gordon11,Baker12}. The basic idea is to treat the fractional deviation from the steady-state concentration as a cumulative distribution for the local accumulation time. One of the useful features of the local accumulation time is that it takes into account the fact that different spatial regions can relax at different rates. (This contrasts with a global measure of the relaxation rate based on the principal nonzero eigenvalue of the negative Laplacian.) In addition, for linear diffusion problems, the mean accumulation time can be calculated by solving the diffusion equation in Laplace space and then taking the small-$s$ limit. This avoids the difficulty in obtaining the full time-dependent solution by evaluating the inverse Laplace transform \cite{Ramanan90}.

The structure of the paper is as follows. In section II we consider a pair of cells coupled by a single static gap junction and give a definition of the local accumulation time. We solve the resulting diffusion equation in Laplace space, and then use this to calculate both the steady-state flux and the corresponding accumulation time. We show that the latter is a monotonically increasing function of spatial position, with a jump discontinuity at the gap junction. This discontinuity vanishes in the limit $\mu \rightarrow \infty$.
In section III we consider the more complicated example of a pair of cells connected via a stochastically-gated gap junction. Defining the steady-state flux and accumulation time in terms of the solution to the first-order moment equations, we obtain analogous results to the static case. Finally, in section IV our analysis is generalized to the case of a linear array of cells with nearest neighbor gap junction coupling. Throughout the paper we fix the length-scale by taking the size of a cell to be $L=1$. The time-scale is then given by $L^2/D$.

\section{A pair of cells coupled by a single gap junction}

\begin{figure}[b!]
\begin{center}
\includegraphics[width=8cm]{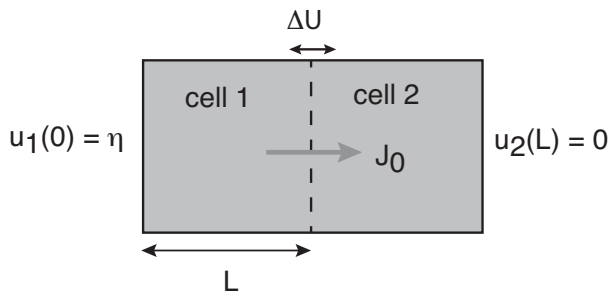}
\caption{Pair of cells coupled by a single static gap junction. At steady-state there is a uniform flux $J_0$ through each cell but a jump discontinuity $\Delta U =-J_0/\mu$ in the concentration across the gap junction, where $\mu$ is the effective channel permeability.}
\label{fig1}
\end{center}
\end{figure}

We begin by considering a pair of identical cells coupled by a single gap junction, as shown in Fig. \ref{fig1}. Following previous studies \cite{Ramanan90,Keener09,Bressloff16}, we treat each cell as a one-dimensional compartment of length $L$ and represent the gap junction as a resistive pore with some permeability $\mu$. For the moment we assume that $\mu$ is given; a mechanism for generating $\mu$ based on stochastic gating will be considered in section 3; see also Ref. \cite{Bressloff16}. Let $u_j(x,t)$, $0<x<L$, denote the concentration of diffusing particles in the $j$-th cell, $j=1,2$. We then have a pair of diffusion equations
\begin{equation}
\label{2udet}
\frac{\partial u_j}{\partial t}=D\frac{\partial^2 u_j}{\partial x^2},\quad x\in (0,L),\, t>0,\ j=1,2.
\end{equation}
These are supplemented by the interior boundary conditions 
\begin{equation}
\label{2bcdet}
-D\frac{\partial u_1(L,t)}{\partial x} =-D\frac{\partial u_2(0,t)}{\partial x} =\mu[u_1(L,t)-u_2(0,t)],
\end{equation}
where $\mu$ is the effective channel permeability, and the exterior boundary conditions
\begin{equation}
 u_1(0,t)=\eta,\quad u_2(L,t)=0.
 \end{equation}
 The interior boundary conditions ensure continuity of flux across the gap junction, which depends on the difference in concentrations on either side of the gap junction. A constant concentration difference is maintained between the exterior boundaries of the two cells. (One way to motivate a 1D model is to assume that the external concentrations are uniform with respect to the other spatial dimensions of the cells, and that there is an array of uniformly distributed gap junctions at the interior boundary that can be reduced to a single effective gap junction \cite{Keener09}.)
 
 It is straightforward to determine the steady-state concentrations $u_j^*(x)$, $j=1,2$, by setting all time derivatives to zero and assuming that there is a constant diffusive flux $J_0$. The flux is then determined self-consistently by solving the resulting boundary value problem \cite{Keener09}. In particular, one finds that
 \begin{equation}
\label{2J0}
J_0=\frac{D\eta \mu }{2\mu L +D} \equiv \frac{D_e\eta }{2L},
\end{equation}
where $D_e$ is an effective diffusion coefficient:
\begin{equation}
\frac{1}{D_e}=\left [\frac{1}{D}+\frac{1}{2\mu L}\right ].
\end{equation}
As highlighted elsewhere \cite{Brink85,Ramanan90}, the effective diffusivity $D_e$ can be compared with experimental measurements of tagged particles such as fluorescent probes. However, experimentalists are typically interested in extracting separate values for $\mu$ and $D$, whereas the permeability and diffusivity are lumped together in the expression for $D_e$. 
As an alternative, one could determine the full time-dependent solution by solving Eq. (\ref{2udet}) in Laplace space and then evaluating the inverse Laplace transform. The resulting solution could then be fit to experimental data \cite{Ramanan90}. In this paper, rather than obtaining the full time-dependent solution, we focus on a particular characterization of the relaxation to steady state known as the local accumulation time. The latter has previously been developed within the context of diffusion-based morphogenesis \cite{Berez10,Berez11,Gordon11,Baker12}, but has more recently been applied to intracellular protein gradient formation \cite{Bressloff19} and to diffusion processes with stochastic resetting \cite{Bressloff21C}. The accumulation time is easier to calculate than the full time-dependent solution, and provides a more compact representation of the relaxation to steady state. 

Consider the initial conditions $u_j(x,0)=\bar{u}_j(x)$ and define
\begin{equation}
\label{accu}
Z_j(x,t)=1-\frac{u_j(x,t)-\bar{u}_j(x)}{u_j^*(x)-\bar{u}_j(x)},\quad j=1,2,
\end{equation}
as the fractional deviation of the concentration in the $j$-th cell from steady state. Assuming that there is no overshooting, $1-Z_j(x,t)$ can be interpreted as the fraction of the steady-state concentration that has accumulated at $x$ by time $t$. It follows that $-\partial_t Z_j(x,t)dt$ is the fraction accumulated in the interval $[t,t+dt]$. The accumulation time $T_j(x)$ at position $x\in (0,L)$ is then defined as \cite{Berez10,Berez11,Gordon11}:
\begin{equation}
\label{accu2}
T_j(x)=\int_0^{\infty} t\left (-\frac{\partial Z_j(x,t)}{\partial t}\right )dt=\int_0^{\infty} Z_j(x,t)dt.
\end{equation}
For more complicated diffusion problems it is more convenient to calculate the accumulation time in Laplace space. Let $\u(x,s)=\int_0^{\infty}\e^{-st}u(x,t)$ etc. Using the identity 
\begin{equation}
\label{ssu}
u_j^*(x)=\lim_{t\rightarrow \infty} u_j(x,t)=\lim_{s\rightarrow 0}s\widetilde{u}_j(x,s)
\end{equation}
and setting $\widetilde{F}_j(x,s)=s\widetilde{u}_j(x,s)-\bar{u}_j(x)$, the Laplace transform of Eq. (\ref{accu}) gives
\[s\widetilde{Z}_j(x,s)=1-\frac{\widetilde{F}_j(x,s)}{\widetilde{F}_j(x)},\quad \widetilde{F}_j(x)=\lim_{s\rightarrow 0}\widetilde{F}_j(x,s)=u_j^*(x)\]
and, hence
\begin{eqnarray}
 T_j(x)&=&\lim_{s\rightarrow 0} \widetilde{Z}_j(x,s) = \lim_{s\rightarrow 0}\frac{1}{s}\left [1-\frac{\widetilde{F}_j(x,s)}{\widetilde{F}_j(x)}\right ] \nonumber \\
 &=&-\frac{1}{u_j^*(x)}
\left .\frac{d}{ds}\widetilde{F}_j(x,s)\right |_{s=0}.
\label{acc}
\end{eqnarray}

\subsection{Solution in Laplace space}

Laplace transforming Eq. (\ref{2udet}) and the associated boundary conditions gives
\begin{equation}
\label{LTu}
\frac{\partial^2 \u_j}{\partial x^2}-k^2\u_j=-D^{-1}\bar{u}_j(x),\quad x\in (0,L),\, t>0
\end{equation}
for $j=1,2$, with $k=\sqrt{s/D}$, the initial conditions $u_j(x,0)=\bar{u}_j(x)$, the interior boundary conditions
\begin{equation}
\label{LTbc}
-D\frac{\partial \u_1(L,s)}{\partial x} =-D\frac{\partial \u_2(0,s)}{\partial x} =\mu[\u_1(L,s)-\u_2(0,s)],
\end{equation}
and the exterior boundary conditions
\begin{equation}
 \u_1(0,s)=\frac{\eta}{s},\quad \u_2(L,s)=0.
 \end{equation}
 For simplicity, we assume that the cells are initially empty so that $\bar{u}_j\equiv 0$. The general solution within each cell then has the form
 \begin{equation}
 \label{gen}
 \u_j(x,s)=A_j\e^{-kx}+B_j\e^{kx},\quad j=1,2.
 \end{equation}
 Substituting these solutions into the various boundary conditions generates four equations for the four unknown coefficients $A_{1,2},B_{1,2}$:
 \begin{subequations}
 \label{cond}
 \begin{align}
 &A_1=\frac{\eta}{s}-B_1,\\
 &A_2=-B_2\e^{2kL},\\
 &A_1\e^{-kL}-B_1\e^{kL}=A_2-B_2,\\
 &A_2-B_2=\frac{\mu}{kD}\left [A_1\e^{-kl}+B_1\e^{kL}-A_2-B_2\right ].
 \end{align}
 \end{subequations}
 Using Eqs. (\ref{cond}a,b) to eliminate $A_1$ and $A_2$ from (\ref{cond}c) implies that
\begin{equation}
B_1=\frac{B_2\left (1+\e^{2kL}\right )+\eta\e^{-kL}/s}{2\cosh kL}.
\end{equation}
Finally, eliminating $B_1$ from Eq. (\ref{cond}d) gives
\begin{equation}
\label{B2}
B_2=-\frac{\mu \eta}{2s}\frac{\e^{-2kL}[1+\tanh kL]}{kD\cosh kL +2\mu \sinh kL}.
\end{equation}
Expressing the solutions (\ref{gen}) in terms of $B_2$, we have
\begin{subequations}
 \label{gen2}
 \begin{align}
 \u_1(x,s)&=\frac{\eta}{s}\left (\e^{-kx}+\e^{-kL}\frac{\sinh kx}{\cosh kL}\right )+2B_2\e^{kL}\sinh kx,\\
 \u_2(x,s)&=-2B_2\e^{kL}\sinh k[L-x].
 \end{align}
 \end{subequations}

The solution in Laplace space can now be used to determine the steady-state concentrations and flux according to Eq. (\ref{ssu}), noting that $k=\sqrt{s/D}\rightarrow 0$ as $s\rightarrow 0$.
We thus find from Eqs. (\ref{B2}) and (\ref{gen2}) that
\begin{subequations}
\begin{align}
u_1^*(x)&=\eta\left (1-  \frac{\mu x}{D+2\mu L}\right ),\\
u_2^*(x)&= \mu \eta\frac{L-x}{D+2\mu L}.
\end{align}
\end{subequations}
We also recover the flux Eq. (\ref{2J0}). Although it is simpler to solve the steady-state equations directly, the advantage of working in Laplace space is that one can determine the accumulation time (and other quantities) without having to solve another boundary value problem. 

\subsection{Accumulation time}

\begin{figure}[b!]
\begin{center}
\includegraphics[width=8.5cm]{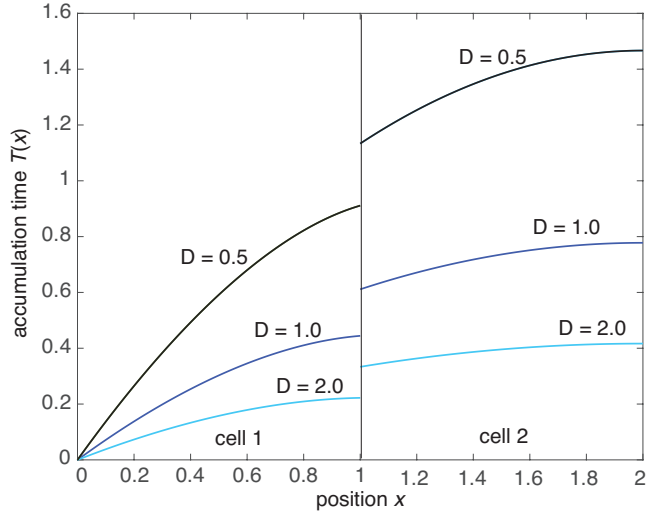}
\caption{Composite accumulation time $T(x)$ for a pair of cells coupled by a static gap junction and various diffusivities $D$. Other parameter values are $\eta =\mu=L=1$.}
\label{fig2}
\end{center}
\end{figure}

The formula (\ref{acc}) implies that we need to evaluate the first derivative of $\widetilde{F}_j(x,s)=s\u_j(x,s)$ in the limit $s\rightarrow 0$. It can be seen from Eqs. \ref{B2}) and (\ref{gen2}) that $\widetilde{F}_j(x,s)$ only depends on $s$ via its $k$-dependence. Therefore, for fixed $x$,
\begin{equation}
\label{Fj}
\frac{d \widetilde{F}_j}{ds}=\frac{dk}{ds}\frac{d \widetilde{F}_j}{dk}=\frac{1}{2\sqrt{sD}}\frac{d \widetilde{F}_j}{dk},\quad j=1,2.
\end{equation}
This will be non-singular in the limit $s\rightarrow 0$ provided that the Taylor expansion of $\widetilde{F}_j$ about $k=0$ is of the form
\begin{equation}
\label{TayF}
\widetilde{F}_j=u^*_j(x)-\frac{k^2}{2} f_j(x)+O(k^3),\quad f_j =-\left .\frac{d^2\widetilde{F}_j}{dk^2}\right |_{k=0}.
\end{equation}
This is indeed found to be the case, see appendix A, so that
\begin{equation}
\left .\frac{d\widetilde{F}_j}{ds}\right |_{s=0}=-\frac{f_j(x)}{2D} \implies T_j(x)=\frac{f_j(x)}{2Du_j^*(x)}.
\end{equation}
Eqs. (\ref{B2}) and (\ref{gen2}a,b) yield the $k$-dependent functions
\begin{align}
\label{F1}
\widetilde{F}_1&=\eta \e^{-kx}+\frac{\eta \e^{-kL}}{\cosh kL}\sinh kx   -\mu \eta \frac{\e^{-kL}[1+\mbox{tanh}kL]\sinh kx}{kD\cosh kL + 2\mu \sinh kL}, 
\end{align}
and
\begin{equation}
\label{F2}
\widetilde{F}_2=\mu \eta \frac{\e^{-kL}[1+\mbox{tanh}kL]\sinh k(L-x)}{kD\cosh kL + 2\mu \sinh kL}.
\end{equation}
Taylor expanding each term in Eqs. (\ref{F1}) and (\ref{F2}) up to $O(k^3)$ allows us to extract the functions $f_j(x)$, $j=1,2$, and thus obtain the following expressions for the local accumulation times, see appendix A:
\begin{align}
T_1(x)&=\frac{\displaystyle 3D+4\mu L}{\displaystyle D+2\mu L}\frac{  L^2}{3D}-\frac{x^2}{6D}-\frac{1}{3D}\left (1-  \frac{\mu x}{D+2\mu L}\right )^{-1}\nonumber \\
&\quad \times \left [x^2-3xL+\frac{\displaystyle 3D+4\mu L}{\displaystyle D+2\mu L}L^2\right ],
\label{resT1}
\end{align}
and
\begin{equation}
\label{resT2}
T_2(x)=\frac{\displaystyle 3D+4\mu L}{\displaystyle D+2\mu L}\frac{  L^2}{3D}-\frac{(L-x)^2}{6D}.
\end{equation}
In contrast to the steady-state flux, the accumulation time could be used to extract values for both $D$ and $\mu$, for example.

\begin{figure}[t!]
\begin{center}
\includegraphics[width=8.5cm]{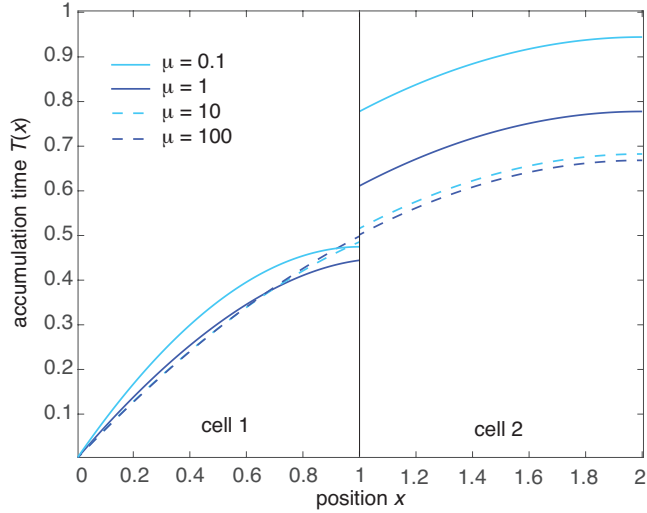}
\caption{Composite accumulation time $T(x)$ for a pair of cells coupled by a static gap junction and various permeabilities $\mu$. Other parameter values are $\eta=D=L=1$.}
\label{fig3}
\end{center}
\end{figure}

Consider the composite accumulation time
\begin{equation}
\label{resT}
T(x)=T_1(x)\Theta(L-x)+T_2(x-L)\Theta(x-L),\quad 0< x <2L,
\end{equation}
where $\Theta(x)$ is the Heaviside function and $T_j(x)$, $j=1,2$, are given by Eqs. (\ref{resT1}) and (\ref{resT2}), respectively. In Fig. \ref{fig2} we show example plots of $T(x)$ for different diffusivities $D$ and $\mu=L=1$. As expected, the accumulation time is a strictly monotonically increasing function of $x$ and a decreasing function of $D$. In addition, we see that there is a jump discontinuity in $T(x)$ at the gap junction connecting the two cells. The corresponding results for various permabilities $\mu$ are shown in Fig. \ref{fig3}. In the limit $\mu\rightarrow \infty$ the gap junction no longer restricts the diffusion of molecules and one simply has a domain of length $2L$. Taking the limit $\mu\rightarrow \infty$ in Eqs. (\ref{resT1}) and (\ref{resT2}) yields the following expression for the composite accumulation time:
\begin{equation}
T(x)=\frac{2L^2}{3D}-\frac{(2L-x)^2}{6D}.
\end{equation}

\setcounter{equation}{0}
\section{Stochastically-gated gap junction}

\begin{figure}[t!]
\centering
\includegraphics[width=6cm]{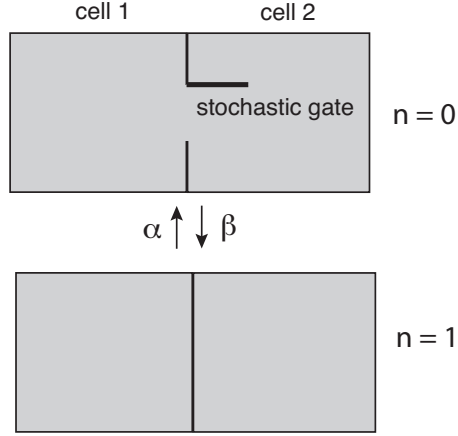}
\caption{Pair of cells coupled by a stochastically-gated (dynamic) gap junction. The gate stochastically switches between an open ($n=0$) and a closed ($n=0$) state according to a two-state Markov process with transition rates $\alpha,\beta$.}\label{fig4}
\end{figure}

We now consider the more complicated problem of a a pair of cells connected by a stochastically-gated (dynamic) gap junction as formulated in Ref. \cite{Bressloff16}, see Fig. \ref{fig4}. The interior boundary between the two cell now randomly switches between an open and a closed state. Let $n(t)$ denote the discrete state of the gate at time $t$ with $n(t)=0$ if the gate is open and $n(t)=1$ if it is closed. 
Assume that transitions between the two states $n=0,1$ are described
by the two-state Markov process, 
\begin{equation*}
0\Markov{\alpha}{\beta}1.
\end{equation*}
The random opening and closing of the gate means that particles diffuse in a random environment according to the piecewise deterministic equations
\begin{equation}
\label{diff}
\frac{\partial u_j}{\partial t}=D\frac{\partial^2u_j}{\partial x^2},\quad x\in (0,L), t>0, \ j=1,2,
\end{equation}
with static exterior boundary conditions
\begin{equation}
u_1(0,t)=\eta>0,\quad u_2(L,t)=0,
\end{equation}
and $n(t)$-dependent boundary conditions on the common interior boundary:
\begin{subequations}
\begin{align}\label{neumannBC}
u_1(L,t)&=u_2(0,t),\,\partial_xu_1(L,t)=\partial_x u_2(0,t) \mbox{ for } n(t)=0,\\
\label{dirichletBC}
&\partial_xu_1(L,t)=0=\partial_x u_2(0,t)\  \mbox{ for } n(t)=1.
\end{align}
\end{subequations}
That is, when the gate is open there is continuity of the concentration and the flux across the common boundary, whereas when the gate is closed the right-hand boundary of cell 1 and the left-hand boundary of cell 2 are reflecting. For simplicity, we assume that the diffusion coefficient is the same in both compartments so that the piecewise nature of the solution is solely due to the switching gate.

Averaging the concentrations with respect to the gate dynamics, we introduce the decompositions
\begin{align}
\label{expU}
U_j(x,t)&\equiv \E[u_j(x,t)] = \E[u_j(x,t)1_{n(t)=0}]+ \E[u_j(x,t)1_{n(t)=1}].
\end{align} 
It can be shown that the components
\begin{equation}
\label{vdirichlet}
V_{j,n}(x,t)=\E[u_j(x,t)1_{n(t)=n}],
\end{equation}
satisfy the first-order moment equations \cite{Bressloff16}
\begin{subequations}
\label{CK}
\begin{eqnarray}
\frac{\partial V_{j,0}}{\partial t} &=&D\frac{\partial^2 V_{j,0}}{\partial x^2}-\beta V_{j,0}+\alpha V_{j,1} \\
 \frac{\partial V_{j,1}}{\partial t} &=& D\frac{\partial^2V_{j,1}}{\partial x^2 }+\beta V_{j,0}-\alpha V_{j,1}
\end{eqnarray}
\end{subequations}
for $x\in (0,L)$ and $j=1,2$, with the exterior boundary conditions
\begin{subequations}
\begin{align}
\label{eBC}
 V_{1,0}(0,t)&=\rho_0 \eta, \quad V_{1,1}(0,t)=\rho_1 \eta,\\
  V_{2,0}(L,t)&= V_{2,1}(L,t)=0, 
\end{align}
and the interior boundary conditions
\begin{align}
\label{iBC}
 V_{1,0}(L,t)&=V_{2,0}(0,t),\, \partial_x V_{1,0}(L,t)=\partial_xV_{2,0}(0,t),\\ \partial_xV_{1,1}(L,t)&=0=\partial_xV_{2,1}(0,t).
\end{align}
\end{subequations}
The boundary condition (\ref{eBC}) follows from noting that if $n(t)=n$ and $x=0$, then $u_1(0,t)=\eta$ with probability one, and thus
\[ V_{1,n}(0,t)=\E[u_1(0,t)1_{n(t)=n}]=\eta\P(n(t)=n)=\eta \rho_n,\]
where
\begin{equation}
\rho_0=\frac{\alpha}{\alpha+\beta},\quad \rho_1=1-\rho_0=\frac{\beta}{\alpha+\beta}.
\end{equation}

Following along similar lines to the static gate, we analyze the diffusion equation in Laplace space. In particular, Eqs. (\ref{CK}a,b) become
\begin{subequations}
\label{CKLT}
\begin{eqnarray}
D\frac{\partial^2 \V_{j,0}}{\partial x^2}-(\alpha +\beta+s) \V_{j,0}+\alpha \U_{j} &=&0,\\
  D\frac{\partial^2 \V_{j,1}}{\partial x^2}-(\alpha +\beta+s) \V_{j,1}+\beta \U_{j} &=&0\end{eqnarray}
\end{subequations}
for $x\in (0,L)$ and $j=1,2$; the boundary conditions are the same. 
From the interior boundary conditions (\ref{iBC}), we set 
\[\partial_x \V_{1,0}(L,s)=\partial_x\V_{2,0}(0,s)=\Gamma(s),\]
with $\Gamma(s)$ to be determined later by imposing $\V_{1,0}(L,s)=\V_{2,0}(0,s)$. Adding equations (\ref{CK}a) and (\ref{CK}b) then gives 
\begin{equation}
\frac{\partial^2 \U_j}{\partial x^2}-k \U_j=0,\quad x\in (0,L),\quad k=\sqrt{s/D},
\end{equation}
with the boundary conditions
\begin{subequations}
\label{bcUj}
\begin{align}
\U_1(0,s)&=\frac{\eta}{s},\quad \partial_x \U_1(L,s)=\Gamma(s), \\
\partial_x\U_2(0,s)&=\Gamma(s),\quad \U_2(L,s)=0.
\end{align}
\end{subequations}
The corresponding solutions are
\begin{subequations}
\label{Uj}
\begin{align}
\U_1(x,s)&=\frac{\eta}{s} \frac{\cosh k(L-x)}{\cosh kL}+\frac{\Gamma}{k} \frac{\sinh kx}{\cosh kL},\\
\U_2(x,s)&=-\frac{\Gamma}{k} \frac{\sinh k(L-x)}{\cosh kL}.
\end{align}
\end{subequations}
We can now substitute for $\U_j(x,s)$ in equation (\ref{CKLT}b), say, and determine the components $\V_{j,1}(x,s)$ for the two cells.

Introduce the Green's functions $G_j(x,y;s)$ according to
\begin{align}
\label{Green}
\frac{\partial^2G_j(x,y;s)}{\partial y^2}-\xi(s)^2 G_j(x,y;s)=-\delta(x-y),
\end{align}
with the boundary conditions
\begin{subequations}
\begin{align}
G_1(x,0;s)&=0,\quad \partial_yG_1(x,L;s)=0, \\
\partial_yG_2(x,0,s)&=0,\quad G_2(x,L;s)=0,
\end{align}
\end{subequations}
and
\begin{equation}
\xi(s)=\sqrt{\frac{\alpha+\beta+s}{D}}.
\end{equation}
The Green's functions have the explicit expressions
\begin{subequations}
\label{Gj}
\begin{align}
G_1(x,y;s)&=\left \{ \begin{array}{cc} A \sinh \xi y \cosh \xi(L-x)& \mbox{ for } y<x\\
 A \sinh \xi x \cosh \xi(L-y) &\mbox{ for } y >x, \end{array}
 \right .\\
G_2(x,y;s)&=\left \{ \begin{array}{cc} A \cosh \xi y \sinh \xi(L-x)& \mbox{ for } y<x\\
 A \cosh \xi x \sinh \xi(L-y) &\mbox{ for } y >x, \end{array}
 \right . 
\end{align}
\end{subequations}
with $A=\xi \cosh \xi L$.
The solutions of  (\ref{CKLT}b) can then be expressed in terms of the Green's functions as follows:
\begin{subequations}
\label{V1j}
\begin{align}
\V_{1,1}(x,s)&=\frac{\beta}{D}\int_0^L G_1(x,y;s)\U_1(y,s)dy  +\frac{\rho_1 \eta}{s} \partial_y G_1(x,0;s),\\
\V_{2,1}(x,s)&=\frac{\beta}{D}\int_0^L G_2(x,y;s)\U_2(y,s)dy.
\end{align}
\end{subequations}
Finally, the unknown function $\Gamma(s)$ is determined by requiring that $\V_{1,0}(L,s)=\V_2(0,s)$ (continuity of the concentration when the gate is open), that is,
\begin{equation}
\label{gamcon}
\U_1(L,s)-\V_{1,1}(L,s)=\U_2(0,s)-\V_{2,1}(0,s),
\end{equation}
with $\U_j$ given by Eqs (\ref{Uj}a,b) and $\V_{j,1}$ given by Eqs. (\ref{V1j}a,b).

\subsection{Steady-state flux} As a check on the above analysis, we begin by deriving the mean steady-state flux $J_0$, which was previously obtained by solving the steady-state first moment equations directly \cite{Bressloff16}. Eqs. (\ref{bcUj}) imply that 
\begin{equation}
J_0=-D\lim_{s\rightarrow 0} s \partial_x\U_1(L,s) =-D\lim_{s\rightarrow 0} s\Gamma(s).
\end{equation}
Multiplying Eqs. (\ref{Uj}a,b) by $s$ and taking the limit $s\rightarrow 0$ gives
\begin{subequations}
\label{limUj}
\begin{align}
U_1^*(x)&=\lim_{s\rightarrow 0}s\U_1(x,s)=\eta -\frac{J_0 x}{D},\\U_2^*(x)&= \lim_{s\rightarrow 0}s\U_2(x,s)=\frac{J_0(L- x)}{D} .
\end{align}
\end{subequations}
Eq. (\ref{gamcon}) thus implies that
\begin{equation}
\label{gamcon2}
\eta -\frac{2J_0L}{D}=\lim_{s\rightarrow 0} s[\V_{1,1}(L,s)-\V_{2,1}(L,s)].
\end{equation}
Next, multiplying Eq. (\ref{V1j}a) by $s$ and taking the limit $s\rightarrow 0$ shows that
\begin{align*}
\lim_{s\rightarrow 0} s \V_{1,1}(x,s)&=\frac{\beta}{D}\int_0^L G_1(x,y;0)\left [\eta -\frac{yJ_0}{D}\right ]dy +\rho_1 \partial_yG_1(x,0;0).
\end{align*}
Multiplying Eq. (\ref{Green}) by $y^m$ for $j=1$, $m=0,1$, and integrating with respect to $y$, we find that
\begin{align*}
\int_0^LG_1(x,y,0)dy&=\frac{1}{\xi_0^2}\left (1-\partial_yG_1(x,0;0)\right ),\\ \int_0^LyG_1(x,y,0)dy&=\frac{1}{\xi_0^2}\left (x-\frac{\sinh \xi_0 x}{\xi_0 \cosh \xi_0 L}\right ),
\end{align*}
with $\xi_0 = \sqrt{(\alpha+\beta)/D}$. We thus obtain the result
\begin{equation}
\label{limV1}
\lim_{s\rightarrow 0} s \V_{1,1}(x,s)=\frac{\rho_1J_0}{\xi_0D} \frac{\sinh \xi_0 x}{\cosh\xi_0 L}+\rho_1\left (\eta -\frac{x J_0}{D} \right ).
\end{equation}
Similarly, 
\begin{align*}
\lim_{s\rightarrow 0} s \V_{2,1}(x,s)&=\frac{\beta}{D}\int_0^L G_2(x,y;0)\frac{J_0(L- y)}{D}dy,
\end{align*}
and
\begin{align*}
&\int_0^LG_2(x,y,0)dy=\frac{1}{\xi_0^2}\left (1+\partial_yG_2(x,L;0)\right ),\\ 
&\int_0^LyG_2(x,y,0)dy =\frac{1}{\xi_0^2}\left (x+\frac{\sinh \xi_0 (L-x)}{\xi_0\cosh \xi_0 L}+L\partial_yG_2(x,L;0)\right ),
\end{align*}
so that
\begin{equation}
\label{limV2}
\lim_{s\rightarrow 0} s \V_{2,1}(x,s)=-\frac{\rho_1J_0}{\xi_0D} \frac{\sinh(\xi_0[L-x])}{\cosh(\xi_0L)}+\frac{\rho_1J_0(L-x)}{D}.
\end{equation}
Finally, substituting Eqs. (\ref{limV1}) and (\ref{limV2}) into (\ref{gamcon2}) and rearranging yields the following expression for the mean steady-state flux:
\begin{equation}
\label{J0switch}
J_0=\frac{D\eta}{2L}\frac{1}{1+(\rho_1/\rho_0)\tanh(\xi_0L)/\xi_0L},
\end{equation}
which recovers the result of Ref. \cite{Bressloff16}. Moreover,
comparison with equation (\ref{2J0}) implies that the stochastically-gated gap junction has the effective permeability $\mu_e$ and diffusivity $D_e$ of the form
\begin{equation}
\label{mum}
\frac{1}{\mu_e}=\frac{2\rho_1}{\rho_0}\frac{\tanh(\xi L)}{\xi D} ,\quad D_e=\frac{2\mu_e LD}{2\mu_eL+D}.
\end{equation}
It is useful to discuss a few limiting cases. First, in the fast switching limit $\xi _0 \rightarrow \infty$, we have $J_0\rightarrow \eta D/2L$, $\mu_e\rightarrow \infty$ and Eq. (\ref{limUj}) reduces to the continuous steady-state solution
\[U_1^*(x)=\eta\left (1-\frac{x}{2L}\right ),\quad U_2^*(x)=\eta \frac{L-x}{2L}.\]
The mean flux through the gate is the same as the steady-state flux without a gate. On the other hand,
for finite switching rates the mean flux $J_0$ is reduced.
In the limit $\alpha \rightarrow 0$ (gate always closed), $J_0\rightarrow 0$ so that $U_1^*(x)=\eta $ for $x\in [0,L)$ and $U_2^*(x)=0$ for $x\in (0,L]$. 

\subsection{Accumulation time}

In the case of a dynamic gate we define the local accumulation time in terms of the first-order moments of the concentration. That is, assuming that the initial concentrations are zero, we take
\begin{equation}
\label{accud}
Z_j(x,t)=1-\frac{U_j(x,t)}{U_j^*(x)},\quad j=1,2,
\end{equation}
where, see Eq. (\ref{expU}),
\begin{equation}
U_j(x,t)\equiv \E[u_j(x,t)],\quad U_j^*(x)=\lim_{t\rightarrow \infty} U_j(x,t).
\end{equation}
The corresponding accumulation times are then defined according to
\begin{eqnarray}
 T_j(x)
=\int_0^{\infty} Z_j(x,t)dt=-\frac{1}{U_j^*(x)}
\left .\frac{d}{ds}\widetilde{F}_j(x,s)\right |_{s=0}
\label{accd}
\end{eqnarray}
with $\widetilde{F}_j(x,s)=s\widetilde{U}_j(x,s)$.
Note that it does not make sense to define an accumulation time for a single realization of the stochastic gate, since one cannot define a steady-state density, that is, $\lim_{t\rightarrow \infty} u_j(x,t)$ does not exist.

Eqs. (\ref{Uj}a) and (\ref{Uj}b) imply that
\begin{align}
\frac{d\widetilde{F}_1}{ds} &=\frac{1}{2\sqrt{sD}}\left [\eta\frac{d}{dk}\frac{\cosh k(L-x)}{\cosh kL}+s\Gamma(s) \frac{d}{dk} \frac{\sinh kx}{k\cosh kL}\right ]+\frac{\sinh kx}{k\cosh kL}\frac{d}{ds}s\Gamma(s),
\end{align}
and
\begin{align}
\frac{d\widetilde{F}_2}{ds} &=-\frac{1}{2\sqrt{sD}} s\Gamma(s) \frac{d}{dk} \frac{\sinh k(L-x)}{k\cosh kL}  -\frac{\sinh k(L-x)}{k\cosh kL} \frac{d}{ds}s\Gamma(s).
\end{align}
In appendix B we show that
\begin{equation}
s\Gamma(s)=-\frac{J_0+sJ_1+O(s^2)}{D}
\end{equation}
for small $s$, with $J_0$ given by Eq. (\ref{J0switch}), so that
\begin{subequations}
\label{Fswitch}
\begin{align}
\left . \frac{d\widetilde{F}_1}{ds}\right |_{s=0} &=\frac{1}{2D}\left (\eta [(L-x)^2-L^2]-\frac{J_0x}{D}[x^2/3-L^2]\right )-\frac{J_1x}{D},
\end{align}
and
\begin{align}
\left .\frac{d\widetilde{F}_2}{ds}\right |_{s=0} &=\frac{J_0(L-x)}{2D^2}[(L-x)^2/3-L^2] +\frac{J_1(L-x)}{D}.
\end{align}
\end{subequations}
The calculation of $J_1$ is presented in appendix B, which yields the expression
\begin{subequations}
\label{J1}
\begin{equation}
J_1=-\frac{\chi J_0L^2}{2D[1+(\rho_1/\rho_0)\tanh \xi_0 L/(\xi_0 L)]},
\end{equation}
where
\begin{align}
\chi&=\frac{1 }{3}+\frac{\rho_1}{\rho_0}\frac{\mbox{tanh}\xi_0 L}{\xi_0L}+\frac{\rho_1}{\rho_0(\xi_0L)^2}\left [1-\frac{\mbox{tanh}\xi_0 L}{\xi_0L}-\mbox{tanh}^2\xi_0 L\right ].
\end{align}
\end{subequations}
In terms of the effective permeability $\mu_{e}$, we have
\begin{equation}
J_1=- \frac{\chi J_0L^2}{D}\frac{\mu_eL}{2\mu_eL+D},
\end{equation}
and
\begin{align}
\chi&= \left (\frac{1}{3}+\frac{\rho_1}{\rho_0\xi_0^2L^2}\right ) +\frac{1}{2}\left (1-\frac{1}{\xi_0^2L^2}\right ) \frac{D}{\mu_eL}
 -\frac{\rho_0}{4\rho_1}\left (\frac{D}{\mu_eL}\right )^2.
\end{align}

\begin{figure}[t!]
\begin{center}
\includegraphics[width=8.5cm]{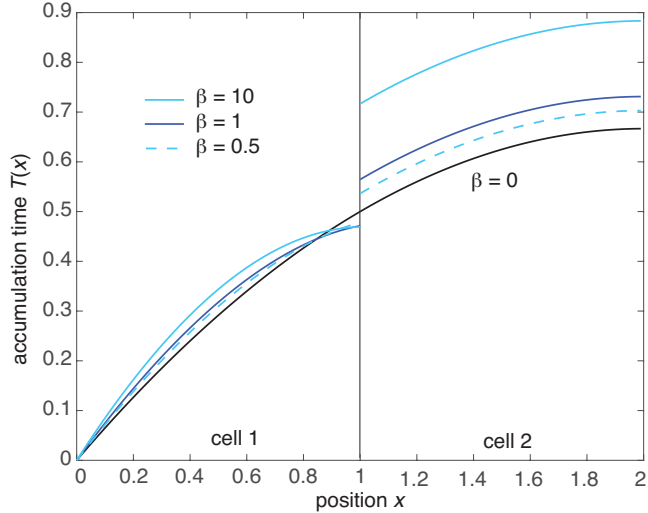}
\caption{Composite accumulation time $T(x)$ for a pair of cells coupled by a dynamic gap junction and various closing rates $\beta$. Other parameter values are $\eta=D=L=\alpha=1$.}
\label{fig5}
\end{center}
\end{figure}

\begin{figure}[t!]
\begin{center}
\includegraphics[width=8.5cm]{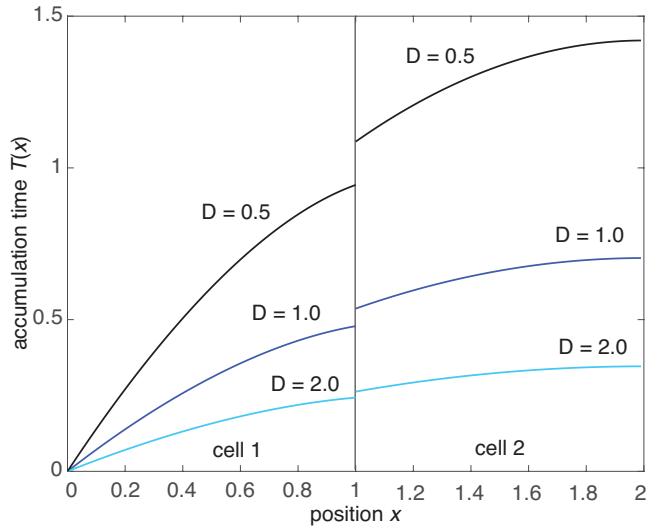}
\caption{Composite accumulation time $T(x)$ for a pair of cells coupled by a dynamics gap junction and various diffusivities $D$. Other parameter values are $\eta=D=L=\alpha=1$ and $\beta=0.5$.}
\label{fig6}
\end{center}
\end{figure}

The accumulation times can now be determined by substituting Eqs. (\ref{limUj}) and (\ref{Fswitch}) into (\ref{accd}) with $J_0$ and $J_1$ given by Eqs. (\ref{J0switch}) and (\ref{J1}), respectively. In Fig. \ref{fig5} we plot the composite accumulation time defined by Eq. (\ref{resT}) as a function of $x\in [0,2L]$ for various values of the closing rate $\beta$. In the limit $\beta \rightarrow 0$ ($\rho_1\rightarrow 0$) the gate is always open, and we recover the continuous accumulation time of a static gap junction in the limit $\mu\rightarrow \infty$, see also Fig. \ref{fig3}. As $\beta$ increases from zero, however, the accumulation time develops a discontinuity at the junction between the two cells, analogous to the static gap junction with finite permeability. Finally, in Fig. \ref{fig6} we show analogous plots for different diffusivities. As expected, increasing $D$ reduces the accummulation time.

\setcounter{equation}{0}
\section{Multi-cell model}

\begin{figure}[t!]
\begin{center}
\includegraphics[width=12cm]{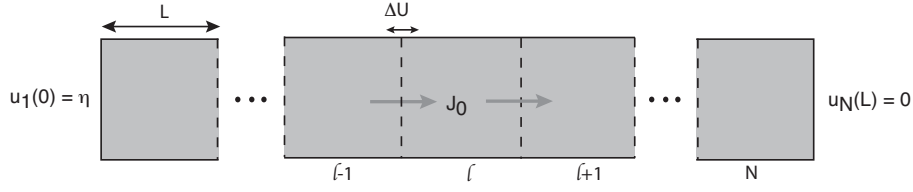}
\caption{One-dimensional array of $N$ cells coupled by gap junctions.}
\label{figm}
\end{center}
\end{figure}

So far we have analyzed the accumulation time for a pair of cells coupled via a single static or dynamic gap junction. We now turn to the case of a 1D array of $N$ identical cells with nearest neighbor gap junction coupling, see Fig. \ref{figm}. We will focus on the static case, since the number of discrete states of $N$ independently gated dynamic junctions grows as $2^N$, which considerably complicates the analysis \cite{Bressloff16}. Suppose that we label the cells by an integer $\ell$, $\ell=1,\ldots,N$, and take the length of each cell to be $L$. Let $u_{\ell}(x,t)$ for $x\in (0,L)$ denote the particle concentration within the interior of the $\ell$-th cell, and consider the diffusion equation
\begin{equation}
\label{mudet}
\frac{\partial u_{\ell}}{\partial t}=D\frac{\partial^2 u_{\ell}}{\partial x^2},\quad x\in (0,L),\, t>0.
\end{equation}
At each of the intercellular boundaries, the concentration is discontinuous due to the permeability of the gap junctions. The generalization of Eq. (\ref{2bcdet}) is
\begin{equation}
\label{mbcdet}
-D\frac{\partial u_{\ell}(L,t)}{\partial x} =-D\frac{\partial u_{\ell+1}(0,t)}{\partial x} =\mu[u_{\ell}(L,t)-u_{\ell+1}(0,t)]
\end{equation}
for $\ell=1,\ldots,N-1$, and the exterior boundary conditions are taken to be
\begin{equation}
 u_1(0,t)=\eta,\quad u_N(L,t)=0.
 \end{equation}

\subsection{Solution in Laplace space}

Laplace transforming Eq. (\ref{mudet}) and the associated boundary conditions gives
\begin{equation}
\label{mLTu}
\frac{\partial^2 \u_{\ell}}{\partial x^2}-k^2\u_{\ell}=-D^{-1}\bar{u}_{\ell}(x),\quad x\in (0,L),\, t>0
\end{equation}
for $\ell=1,\ldots,N$
with the boundary conditions
\begin{equation}
\label{mLTbc}
-D\frac{\partial \u_{\ell}(L,s)}{\partial x} =-D\frac{\partial \u_{\ell+1}(0,s)}{\partial x} =\mu[\u_{\ell}(L,s)-\u_{\ell+1}(0,s)]
\end{equation}
for $\ell=1,\ldots,N-1$, and
\begin{equation}
 \u_1(0,s)=\frac{\eta}{s},\quad \u_N(L,s)=0.
 \end{equation}
(We again assume that the cells are initially empty.) The general solution within each cell then has the form
 \begin{equation}
 \label{mgen}
 \u_{\ell}(x,s)=A_{\ell}\e^{-kx}+B_{\ell}\e^{kx},\quad \ell=1,\ldots,N.
 \end{equation}
 Substituting these solutions into the various boundary conditions generates the following hierarchy:
 \begin{subequations}
 \label{mcond}
 \begin{align}
 &A_1=\frac{\eta}{s}-B_1,\\
 &A_{\ell}\e^{-kL}-B_{\ell}\e^{kL}=A_{\ell+1}-B_{\ell+1},\\
 &A_{\ell+1}-B_{\ell+1}=\frac{\mu}{kD}\left [A_{\ell}\e^{-kL}+B_{\ell}\e^{kL}-A_{\ell+1}-B_{\ell+1}\right ],\\
 &A_N=-B_N\e^{2kL}
 \end{align}
 \end{subequations}
Eqs. (\ref{mcond}b,c) hold for $\ell=1,\ldots,N-1$.
 
 Following along analogous lines to Ref. \cite{Ramanan90}, we rewrite Eqs. (\ref{mcond}b,c) as the matrix equation
 \begin{equation}
   \label{matrix}
 \left ( \begin{array}{cc} A_{\ell +1} \\ B_{\ell+1}\end{array}
 \right )={\bf M}(k)\left ( \begin{array}{cc} A_{\ell } \\ B_{\ell}\end{array}
 \right )
 \end{equation}
 for $\ell =1,\dots N-1$ with
  \begin{equation}
 {\bf M}(k)=\left (\begin{array}{cc} \left (1-\frac{\displaystyle kD}{\displaystyle 2\mu}\right )\e^{-kL}& \frac{\displaystyle kD}{\displaystyle 2\mu}\e^{kL}\\ & \\
 -\frac{\displaystyle kD}{\displaystyle 2\mu}\e^{-kL} & \left (1+\frac{\displaystyle kD}{\displaystyle 2\mu}\right )\e^{kL} \end{array}
 \right ). \end{equation}
Let $(\lambda_{\pm},{\bf v}_{\pm})$ denote the eigenpairs of the non-symmetric matrix ${\bf M}$ and set
\begin{equation}
\label{ellie}
\left ( \begin{array}{cc} A_{\ell } \\ B_{\ell}\end{array}
 \right )=a_{\ell}{\bf v}_++b_{\ell} {\bf v}_-.
 \end{equation}
 (For ease of notation, we drop the explicit dependence on $k=\sqrt{s/D}$.)
 Substituting the eigenfunction expansion into the matrix equation and iterating shows that
 \begin{equation}
 \label{ab}
 a_{\ell}=\lambda_+^{\ell-1}a_1,\quad b_{\ell}=\lambda_-^{\ell-1}b_1.
 \end{equation}
In addition, Eqs. (\ref{mcond}a,d) imply that
\begin{subequations}
\label{a1aN}
\begin{align}
a_1&=\frac{\eta}{s} \Lambda_+-B_1  \widehat{\Lambda}_+ ,\
  b_1=\frac{\eta}{s} \Lambda_- -B_1  \widehat{\Lambda}_-,\\
a_N&=B_N\overline{\Lambda}_+ ,\quad b_N=B_N\overline{\Lambda}_-, 
 \end{align}
 \end{subequations}
 where
 \begin{align}
 \Lambda_{\pm}&=\widetilde{\bf v}_{\pm }^{\top} \cdot \left ( \begin{array}{cc} 1 \\ 0\end{array}\right ),\quad  \widehat{\Lambda}_{\pm}=\widetilde{\bf v}_{\pm}^{\top} \cdot \left ( \begin{array}{cc} 1 \\ -1\end{array}\right ),\nonumber \\
  \overline{\Lambda}_{\pm}&=\widetilde{\bf v}_{\pm}^{\top} \cdot \left ( \begin{array}{cc} -\e^{2kL} \\ 1\end{array}
 \right )=-\widehat{\Lambda}_{\pm} -\left (\e^{2kL}-1\right ) \Lambda_{\pm}.
 \label{lam}
 \end{align}
 We have introduced the dual vectors $\widetilde{\bf v}_j$ such that
 \begin{equation}
 \widetilde{\bf v}_i^{\top} \cdot {\bf v}_j=\delta_{i,j}.
 \end{equation}
 
 Therefore, setting $\ell=N$ in Eq. (\ref{ab}) yields a pair of equations for the two remaining unknowns $B_1$ and $B_N$:
 \begin{subequations}
 \begin{align}
 B_N\overline{\Lambda}_+&=\lambda_+^{N-1}\left [\frac{\eta}{s} \Lambda_+-B_1  \widehat{\Lambda}_+\right ],\\
 B_N\overline{\Lambda}_-&=\lambda_-^{N-1}\left [\frac{\eta}{s} \Lambda_--B_1  \widehat{\Lambda}_-\right ] .\end{align}
 \end{subequations}
Eliminating the coefficient $B_N$ we find after some algebra that $B_1$ has the solution
\begin{equation}
B_1=g(k)\frac{\eta}{s},
\end{equation}
with
\begin{align}
\label{B1f}
g(k)&= \frac{\displaystyle \frac{\overline{\Lambda}_-}{\overline{\Lambda}_+}\left (\frac{\lambda_+}{\lambda_-}\right )^{N-1}\Lambda_+-\Lambda_-}{\displaystyle \frac{\overline{\Lambda}_-}{\overline{\Lambda}_+}\left (\frac{\lambda_+}{\lambda_-}\right )^{N-1}\widehat{\Lambda}_+-\widehat{\Lambda}_-}=  \frac{\displaystyle \frac{1 +\left (\e^{2kL}-1\right ) \Sigma_{-}}{1+\left (\e^{2kL}-1\right ) \Sigma_{+}}\left (\frac{\lambda_+}{\lambda_-}\right )^{N-1}\Sigma_+-\Sigma_-}{\displaystyle \frac{1+\left (\e^{2kL}-1\right ) \Sigma_{-}}{1+\left (\e^{2kL}-1\right ) \Sigma_{+}}\left (\frac{\lambda_+}{\lambda_-}\right )^{N-1} -1} \end{align}
and
\begin{equation}
\Sigma_{\pm}=\frac{\Lambda_{\pm}}{\widehat{\Lambda}_{\pm}}.
\end{equation}
(Recall that the eigenpairs $(\lambda_{\pm},{\bf v}_{\pm})$ depend on $k$.)

Having found $B_1$, all of the coefficients $\{A_{\ell},B_{\ell},\ell=1,\ldots,N\}$ are then determined from Eqs. (\ref{ellie}), (\ref{ab}) and (\ref{a1aN}a).
In addition, we can now calculate the steady-state concentrations using
\begin{align}
U_{\ell}^*(x)=\lim_{s\rightarrow 0}s\u_{\ell}(x,s),
\end{align}
and
\begin{align}
\label{uell}
\u_{\ell}(x,s)={\bf f}(x) \cdot \left ( \begin{array}{cc} A_{\ell } \\ B_{\ell}\end{array}
 \right )=a_{\ell}{\bf f}(x) \cdot {\bf v}_++b_{\ell} {\bf f}(x)^{\top}\cdot {\bf v}_-
 \end{align}
 where
$ {\bf f}(x)=  (  \e^{-kx} ,\e^{kx} )^{\top}
 $.
 In particular,
 \begin{equation}
 U_1^*(x)=\eta \lim_{s\rightarrow 0} \left [\e^{-kx}+2g(k) \sinh kx\right ],
 \end{equation}
and the corresponding steady-state flux is
 \begin{equation}
 J_0=-\eta D\lim_{s\rightarrow 0} \left [-k   \e^{-kx}+2kg(k) \cosh kx\right ].
 \end{equation}
 
 It remains to calculate the eigenfunctions and eigenvalues of the matrix ${\bf M}(k)$ and to determine the function $g(k)$. We find that
 \begin{subequations}
 \label{eigp}
 \begin{align}
 \lambda_{\pm}&=\cosh kL+\frac{kD}{2\mu} \sinh kL  \pm \sqrt{\left (\cosh kL+\frac{kD}{2\mu} \sinh kL\right )^2-1},\  \\
 {\bf v}_{\pm}&=\left (\begin{array}{c}\left [ 1+\frac{\displaystyle kD}{\displaystyle 2\mu}\right ] \e^{kL}-\lambda_{\pm} \\  \\ \frac{\displaystyle kD}{\displaystyle 2\mu}\e^{-kL} \end{array}
 \right ),\quad
 \widetilde{\bf v}_{\pm}={\mathcal N}_{\pm}\left (\begin{array}{c}\left [ 1-\frac{\displaystyle kD}{\displaystyle 2\mu}\right ] \e^{-kL}-\lambda_{\mp} \\ \\ \frac{\displaystyle kD}{\displaystyle 2\mu}\e^{kL} \end{array}
 \right ).
 \end{align}
 \end{subequations}
 The normalization factors ${\mathcal N}_{\pm}$ ensure the conditions $\widetilde{\bf v}_{\pm}\cdot {\bf v}_{\pm}=1$. However, they cancel out in Eq. (\ref{B1f}). One important observation is that 
 \[\lambda_{\pm}=1\pm k\lambda_{0} +\frac{k^2}{2}\lambda_0^2+O(k^3),\quad \widetilde{\bf v}_{\pm}=k{\bf v}^{(1)}_{\pm}+O(k^2),\]
 with
 \begin{align}
 \label{lam2}
 \lambda_0&=\sqrt{L^2+DL/\mu},\quad
 \widetilde{\bf v}_{\pm}^{(1)}=\left (\begin{array}{c}  -L-\frac{\displaystyle D}{\displaystyle 2\mu}\pm \lambda_0 \\ \\ \frac{\displaystyle D}{\displaystyle 2\mu}\end{array}
 \right ).
 \end{align}
It also follows from Eq. (\ref{lam}) that $\Sigma_{\pm}(k)= \Sigma_{\pm}(0)+O(k)$ with
 \begin{equation}
 \Sigma_{\pm}(0)=\frac{ L+\frac{\displaystyle D}{\displaystyle 2\mu}\mp\lambda_{0}}
 { L+\frac{\displaystyle D}{\displaystyle \mu}\mp \lambda_{0}}+O(k).
 \end{equation}
 Moreover,
 \begin{align}
 \Sigma_-(0)-\Sigma_+(0)&=\frac{D}{\mu}\frac{\lambda_0}{[L+D/\mu-\lambda_0][L+D/\mu+\lambda_0]} = \frac{\lambda_0}{L+D/\mu}.
 \label{sig2}
 \end{align}
 
The above analysis implies that $g(k)=g_{-1}/k+O(1)$ and thus
 \begin{equation}
 J_0=-2\eta Dg_{-1}.
 \end{equation}
Finally, we can determine $g_{-1}$ by substituting Eqs. (\ref{lam2}) and (\ref{sig2}) into Eq. (\ref{B1f}):
 \begin{align}
 g_{-1}&= \frac{\displaystyle \Sigma_+(0)-\Sigma_-(0)}{\displaystyle 2L[ \Sigma_{-}(0)-\Sigma_+(0)]+2(N-1)\lambda_0}\nonumber\\
 &=-\frac{1}{2}\frac{1}{L+(N-1)[L+D/\mu ]}.
 \end{align}
 We thus recover the classical result for the steady-state flux through the coupled system shown in Fig. \ref{figm} \cite{Ramanan90,Keener09,Bressloff16}:
\begin{equation}
J_0=\frac{D\eta\mu }{N\mu L+(N-1)D}.
\end{equation}
Introducing the effective diffusion coefficient $D_e$ according to
\begin{equation}
J_0=\frac{D_e\eta}{NL},
\end{equation}
we see that for large $N$
\begin{equation}
\frac{1}{D_e}=\left [\frac{1}{D}+\frac{1}{\mu L}\right ].
\end{equation}

\subsection{Accumulation time}

\begin{figure}[t!]
\begin{center}
\includegraphics[width=8cm]{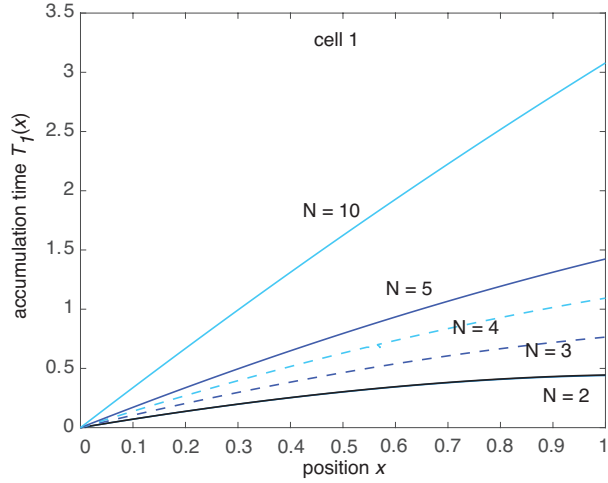}
\caption{Accumulation time $T_1(x)$ in the first cell of the linear array shown in Fig. \ref{figm} for different values of $N$. Other parameter values are $D=\mu=\eta=\mu =1$.}
\label{fig8}
\end{center}
\end{figure}

Deriving an explicit expression for the accumulation time is nontrivial for general $N$. For the sake of illustration, consider the accumulation time $T_1(x)$, $x\in [0,L]$, in the first cell. In this case,
\begin{equation}
\label{NT1}
T_1(x)=-\frac{\eta}{U_1^*(x)}\frac{1}{2D} \lim_{k\rightarrow 0}\frac{1}{k}\frac{d}{dk} \left [\e^{-kx}+2g(k) \sinh kx\right ]
\end{equation}
with $g(k) $ given by Eq. (\ref{B1f}). In Fig. \ref{fig8} we plot $T_1(x)$ as a function of $x$ for various cell numbers $N$. The limit on the right-hand side of Eq. (\ref{NT1}) is obtained by substituting for $g(k)$ using Eq. (\ref{B1f}) and numerically evaluating the resulting derivative numerically. As a useful check, we note that the numerical plot converges to the analytical result (\ref{resT1}) when $N=2$. It can be seen that increasing the number of cells in the array increases the accumulation time in the first cell. 

\begin{figure}[b!]
\begin{center}
\includegraphics[width=8cm]{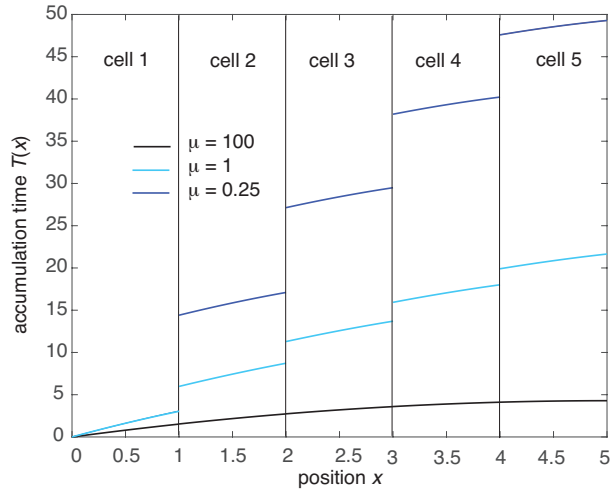}
\caption{Composite accumulation time $T(x)$ of the first five cells in a linear array of size $N=10$ and various values of the permeability $\mu$. Other parameter values are $L=D=\eta=1$.. Other parameter values are $D=\mu=\eta=\mu =1$.}
\label{fig9}
\end{center}
\end{figure}

In order to determine the corresponding accumulation time $T_{\ell}(x)$ in the $\ell$-th cell, we use Eqs. (\ref{ab}), (\ref{a1aN}a) and (\ref{uell}). That is, setting $F_{\pm}(x)={\bf f}(x)^{\top}\cdot {\bf v}_{\pm}$,
\begin{align}
s\u_{\ell}(x,s)&=s\left (a_{\ell}F_+(x)+b_{\ell}F_-(x)\right )\nonumber \\
&=s\left (\lambda_+^{\ell-1}a_1F_+(x)+\lambda_-^{\ell-1}b_1F_-(x)\right )\nonumber \\
&=\eta \left ( \Lambda_+-g(k)  \widehat{\Lambda}_+ \right )\lambda_+^{\ell-1}F_+(x)  +\eta \left (  \Lambda_--g(k) \widehat{\Lambda}_-\right )\lambda_-^{\ell-1}F_-(x)\nonumber \\
&\equiv F_{\ell}(x,k).
\end{align}
We have used the result $B_1=\eta g(k)/s$. Note that $\lambda_{\pm}, \Lambda_{\pm}$ and $F_{\pm}(x)$ are also functions of $k$, which means that $s\u_{\ell}(x,s)$ only depends on $s$ via its dependence on $k$. The generalization of Eq. (\ref{NT1}) is thus
\begin{equation}
\label{NTell}
T_{\ell}(x)=-\frac{\eta}{U_{\ell}^*(x)}\frac{1}{2D} \lim_{k\rightarrow 0}\frac{1}{k}\frac{d}{dk} F_{\ell}(x,k).
\end{equation}
Define the composite accumulation time
\begin{equation}
T(x) =\sum_{\ell=1}^NT_{\ell}(x-(\ell-1)L)\chi_{\ell}(x),\quad 0 < x <NL,
\end{equation}
where $\chi_{\ell}(x)=1$ for $x\in [(\ell-1)L,\ell L)$ and is zero otherwise. In Fig. \ref{fig9} we plot $T(x)$ for the first five cells in a linear array of size $N=10$ and various permeabilities $\mu$. As in the case of two cells, see Fig. \ref{fig3}, $T(x)$ is a monotonically increasing function of $x$ with jump discontinuities at the gap junctions. In the limit $\mu\rightarrow \infty$ the discontinuities vanish and $T(x)$ becomes a continuous function of $x$.


\section{Discussion}

In this paper we calculated the local accumulation time for intracellular diffusion in cells with gap-junction coupling. We considered a pair of cells connected by either a static or a dynamic gap junction. In both cases, we showed that the accumulation time is a monotonically increasing function of spatial position with a jump discontinuity at the gap junction. This discontinuity vanished in the limit $\mu \rightarrow \infty$ for a static junction with permeability $\mu$ and in the limit $\beta \rightarrow 0$ for a stochastically-gated junction with rate of closing $\beta$. We also extended our analysis of static gap junctions
 to the case of a linear array of cells with nearest neighbor gap junction coupling. In contrast to the expressions for the steady-state flux, the accumulation times did not simply depend on the effective permeability and cytoplasmic diffusivity via a lumped parameter given by an effective diffusivity.
 
One limitation of our analysis of a stochastically gated gap junction in section III is that we only considered the first-order moment equations for the mean stochastic concentrations $\E[u_j(x,t)]$. Mathematically speaking, one can view diffusion in a randomly switching environment such as a stochastically gated gap junction as an example of a piecewise deterministic partial differential equation (PDE). Previously we have shown how one can analyze such a system by discretizing space and constructing the Chapman-Kolmogorov (CK) equation for the resulting finite-dimensional system \cite{Bressloff15,Bressloff16}. The CK equation can then be used to generate a hierarchy of equations for the $r$-th order moments of the stochastic concentration, which take the form of $r$-dimensional parabolic PDEs in the continuum limit. Although the diffusing particles are non-interacting, statistical correlations arise at the population level due to the fact that they all move in the same randomly switching environment. That is, for the $j$-th cell $\E[u_j(x,t)u_j(y,t)]\neq \E[u_j(x,t)] \E[u_j(y,t)]$.  
 
Another simplification of our analysis was to focus on one-dimensional diffusion models. One natural extension of our work would be to consider higher spatial dimensions and more general geometric configurations of cells. In the case of static gap junctions, Keener and Sneyd \cite{Keener09} have analyzed the steady-state flux for a line of two-dimensional cells with gap-junctional openings in the connecting edges. Using symmetry arguments, they showed how the gap junctions along an edge can be lumped into a single effective junction at the center of each edge, whose relative width characterized the degree of clustering of the gap junctions. In particular, they found that clustered gap junctions lead to a much smaller effective diffusion coefficient $D_e$ for given gap junctional permeability $\mu$. In future work it would be interesting to investigate how such clustering affects the corresponding local accumulation times. An alternative generalization of one-dimensional diffusion would be to consider diffusion on a tree-like structure with stochastically-gated nodes \cite{Bressloff16a}. A number of
biological systems employ branched tree structures in order to distribute nutrients from a single source to many destinations or to gather nutrients from many sources. Examples include plant roots, river basins, neuronal dendrites, and cardiovascular and tracheal systems.

\setcounter{equation}{0}
\renewcommand{\theequation}{A.\arabic{equation}}
\section*{Appendix A: Calculation of accumulation time for a static gate}

Carrying out the Taylor expansions of Eqs. \ref{F1}) and (\ref{F2}) up to $O(k^3)$  we find that
\begin{align*}
\widetilde{F}_1&=\eta [1+(kx)^2/2-k^2xL] -\frac{\mu \eta x}{D+2\mu L} \frac{[1-(kL)^2/2][1+k^2x^2/6]} {1+\frac{\displaystyle 3D+2\mu L}{\displaystyle D+2\mu L}\frac{\displaystyle(kL)^2}{\displaystyle 6}}+O(k^3)\\
&\quad =\eta [1+(kx)^2/2-k^2xL]-[\eta-u_1^*(x)]\left [1+\frac{k^2x^2}{6}-\frac{\displaystyle 3D+4\mu L}{\displaystyle D+2\mu L}\frac{\displaystyle(kL)^2}{\displaystyle 3}\right ] \\&\qquad +O(k^3)\\
&=u_1^*(x)\left [1+\frac{k^2x^2}{6}-\frac{\displaystyle 3D+4\mu L}{\displaystyle D+2\mu L}\frac{\displaystyle(kL)^2}{\displaystyle 3}\right ]+\eta \frac{k^2}{3}\left [x^2-3xL+\frac{\displaystyle 3D+4\mu L}{\displaystyle D+2\mu L}L^2\right ]\\
&\qquad +O(k^3).
\end{align*}
and
\begin{align*}
\widetilde{F}_2&=\mu \eta \frac{\left [1-kL+(kL)^2/2-(kL)^3/6\right ]\left [1+kL-(kL)^3/3\right][k(L-x)+k^3(L-x)^3/6]+\ldots}
{kD[1+(kL)^2/2]+2\mu [kL+(kL)^3/6]+\ldots}\\
&=\frac{\mu \eta(L-x)}{D+2\mu L} \frac{[1-(kL)^2/2][1+k^2(L-x)^2/6]} {1+\frac{\displaystyle 3D+2\mu L}{\displaystyle D+2\mu L}\frac{\displaystyle(kL)^2}{\displaystyle 6}}+O(k^3)\\
&\quad =u_2^*(x)\left [1-\frac{(kL)^2}{2}+\frac{k^2(L-x)^2}{6}-\frac{\displaystyle 3D+2\mu L}{\displaystyle D+2\mu L}\frac{\displaystyle(kL)^2}{\displaystyle 6}\right ]+O(k^3).
\end{align*}
It immediately follows that the Taylor expansions have the general form given by Eq. (\ref{TayF}), namely,
\begin{equation*}
\widetilde{F}_j=u^*_j(x)-\frac{k^2}{2} f_j(x)+O(k^3),\quad f_j =\left .\frac{d^2\widetilde{F}_j}{dk^2}\right |_{k=0}.
\end{equation*}
Hence, $T_j(x)={f_j(x)}/{2Du_j^*(x)}$ and we obtain Eqs. (\ref{resT1}) and (\ref{resT2}).

\setcounter{equation}{0}
\renewcommand{\theequation}{B.\arabic{equation}}
\section*{Appendix B: Calculation of accumulation time for a dynamic gate}

In this appendix we calculate the $O(s)$ contribution to $s\Gamma(s)$, which is needed in order to determine the local accumulation times in the case of a dynamic gate. Substituting Eqs. (\ref{Uj}a,b) and (\ref{V1j}a,b) into (\ref{gamcon}) and rearranging yields
\begin{equation}
\label{sgam}
s\Gamma(s)=-\frac{A(s)}{B(s)},
\end{equation}
where
\begin{align}
A(s)&=\eta -\frac{\beta \eta}{D}\int_0^LG_1(L,y;s)\cosh k(L-y)dy\-\rho_1\eta \cosh( kL)\partial_yG_1(L,0;s) ,
\end{align}
and
\begin{align}
B(s)&=\frac{2\sinh kL}{k}-\frac{\beta}{Dk}\int_0^LG_2(0,y;s)\sinh k(L-y)dy\nonumber \\
&\quad -\frac{\beta}{Dk}\int_0^LG_1(L,y;s)\sinh ky \ dy .
\end{align}
Taylor expanding $A(s)$ and $B(s) $ with respect to $s$, we find that
\begin{equation}
A(s)=A_0+sA_1+O(s^2),\quad B(s)=B_1+sB_1+O(s^2),
\end{equation}
where
\begin{align}
A_0=\rho_0\eta,\quad B_0=2L[\rho_0+\rho_1\tanh \xi_0 L/(\xi_0 L)].
\end{align}
The $O(s)$ coefficients are
\begin{align}
A_1&=-\frac{\beta \eta}{2D^2}\int_0^LG_1(L,y;0)(L-y)^2dy\\\
&\quad -\frac{\beta \eta}{D}\int_0^L\partial_sG_1(L,y;0)dy-\rho_1\eta \partial_s\partial_y G_1(L,0;0) -\rho_1 \eta \frac{L^2}{2D}\partial_yG_1(L,0;0),\nonumber 
\end{align}
and
\begin{align}
&B_1=\frac{L^3}{3D}-\frac{\beta}{6D^2}\int_0^LG_2(0,y;0)(L-y)^3dy -\frac{\beta}{D}\int_0^L\partial_sG_2(0,y;0)(L-y)dy \nonumber \\
&\quad -\frac{\beta}{6D^2}\int_0^LG_1(L,y;0)y^3dy -\frac{\beta}{D}\int_0^L\partial_sG_1(L,y;0)ydy.\
\end{align}
The various Green's function moment can be determined from Eq. (\ref{Green}). First, multiplying Eq. (\ref{Green}) by $(L-y)^2$ and $y^3$, respectively,  for $j=1$ and integrating with respect to $y$, we have
\begin{align*}
&\xi_0^2\int_0^LG_1(L,y;0)(L-y)^2dy =-\left (\frac{2}{\xi_0^2}+L^2\right )\partial_yG_1(L,0;0)+\frac{2}{\xi_0^2},
\end{align*}
and
\begin{align*}
\xi_0^2\int_0^LG_1(L,y;0)y^3dy&=L^3-3L^2G_1(L,L;0) +\frac{6L}{\xi_0^2}\left (1-\frac{\mbox{tanh} \xi_0 L}{\xi_0 L}\right ).
\end{align*}
Similarly, multiplying Eq. (\ref{Green}) by $(L-y)^3$ for $j=2$ and integrating with respect to $y$,
\begin{align*}
&\xi_0^2\int_0^LG_2(0,y,0)(L-y)^3dy=L^3-3L^2 G_2(0,0;0)+\frac{6L}{\xi_0^2}\left (1-\frac{\mbox{tanh} \xi_0 L}{\xi_0 L}\right ).
\end{align*}

Next, introduce the functions
\begin{equation}
H_j(x,y;s)=\partial_sG_j(x,y;s).
\end{equation}
Differentiating both sides of Eq. (\ref{Green}) with respect to $s$ shows that $H_j$ satisfies the inhomogeneous equation
\begin{equation}
\label{GreenH}
\frac{\partial^2H_j(x,y;s)}{\partial y^2}-\xi(s)^2 H_j(x,y;s)=\frac{G_j(x,y;s)}{D}.
\end{equation}
together with the same boundary conditions as $G_j$. It follows that $H_j$ has the solution
\begin{equation}
H_j(x,y;s)=-\frac{1}{D}\int_0^L G_j(x,z;s) G_j(x,z;s)dz.
\end{equation}
Equation (\ref{GreenH}) can now be used to calculate the required moments of $\partial_sG_j$. 
First, multiplying Eq. (\ref{GreenH}) by $y^m$ for $j=1$, $m=0,1$, and integrating gives
\begin{align*}
\xi_0^2\int_0^LH_1(L,y,0)dy&=-\frac{1}{\xi_0^2D}\left (1-\partial_yG_1(L,0;0)\right )-\partial_y\partial_sG_1(L,0;0),\\ \xi_0^2 \int_0^LyH_1(x,y,0)dy&=-\frac{L}{\xi_0^2D}\left (1-\frac{\mbox{tanh} \xi_0 L}{\xi_0 L}\right )-\partial_sG_1(L,L;0).
\end{align*}
Second, multiplying Eq. (\ref{GreenH}) by $L-y$ for $j=2$ and integrating gives
\begin{align*}
\xi_0^2\int_0^LH_2(L,y,0)(L-y)dy&=-\frac{L}{\xi_0^2D}\left (1-\frac{\mbox{tanh} \xi_0 L}{\xi_0 L}\right )-\partial_sG_2(0,0;0).
\end{align*}

Combining all of our results, we find that
\begin{align}
A_1&=-\frac{\rho_1 \eta}{2D}\left [  -\left (\frac{2}{\xi_0^2}+L^2\right )\partial_yG_1(L,0;0)+\frac{2}{\xi_0^2}\right ]\nonumber \\
&\quad + {\rho_1 \eta} \left [\frac{1}{\xi_0^2D}\left (1-\partial_yG_1(L,0;0)\right )+\partial_y\partial_sG_1(L,0;0),\right ]\nonumber \\
&\quad -\rho_1\eta \partial_s\partial_y G_1(L,0;0)-\rho_1 \eta \frac{L^2}{2D}\partial_yG_1(L,0;0)=0 ,
\end{align}
and
\begin{align}
&B_1=\frac{L^3}{3D}-\frac{\rho_1}{6D}\left [L^3 -3L^2 G_2(0,0;0)+\frac{6L}{\xi_0^2}\left (1-\frac{\mbox{tanh} \xi_0 L}{\xi_0 L}\right )\right ]\nonumber \\
&\quad +\rho_1\left [\frac{L}{\xi_0^2D}\left (1-\frac{\mbox{tanh} \xi_0 L}{\xi_0 L}\right ) +\partial_sG_2(0,0;0)\right ]\nonumber \\
&\quad -\frac{\rho_1}{6D}\left [L^3-3L^2G_1(L,L;0) +\frac{6L}{\xi_0^2}\left (1-\frac{\mbox{tanh} \xi_0 L}{\xi_0 L}\right )\right ] \nonumber \\
&\quad +\rho_1 \left [\frac{L}{\xi_0^2D}\left (1-\frac{\mbox{tanh} \xi_0 L}{\xi_0 L}\right )+\partial_sG_1(L,L;0)\right ]\nonumber \\
&=\frac{2L^3-2\rho_1 L^3}{6D}+\frac{\rho_1L^2}{2D}[G_1(L,L;0)+G_2(0,0;0)]\nonumber \\
&\quad +\rho_1[\partial_sG_1(L,L;0)+\partial_sG_2(0,0;0)] \nonumber \\
&=\frac{\rho_0L^3}{3D}+\frac{\rho_1L^3}{D}\frac{\mbox{tanh}\xi_0 L}{\xi_0L}+\frac{\rho_1L}{D\xi_0^2}\left [1-\frac{\mbox{tanh}\xi_0 L}{\xi_0L}-\mbox{tanh}^2\xi_0 L\right ].
\end{align}

Finally, substituting the $s$-expansions of $A(s)$ and $B(s)$ into Eq. (\ref{sgam}, we see that
\begin{align}
s\Gamma(s)&=-\frac{A_0+A_1s+\ldots}{B_0+B_1s +\ldots }=-\frac{A_0}{B_0} +-\frac{A_0B_1}{B_0^2}s+O(s^2), 
\end{align}
which implies that $s\Gamma(s)=-(J_0+J_1s)/D+O(s^2)$ with
\begin{equation}
J_0= \frac{A_0D}{B_0}=\frac{D\eta}{2L}\frac{1}{1+(\rho_1/\rho_0)\tanh(\xi_0L)/\xi_0L},
\end{equation}
which recovers Eq. (\ref{J0switch}),
and
\begin{equation}
J_1=  -A_0B_1\frac{D}{B_0^2}= -\frac{J_0B_1}{B_0}.
\end{equation}


\end{document}